\documentclass[12pt]{article}

\usepackage{a4}
\usepackage{epsfig}

\def\Journal#1#2#3#4{{#1} {\bf #2} (#3) #4.}
\def\etal{{\it et\ al.}}

\def\NPB{{\em Nucl. Phys.} B}

\def\PRD{{\em Phys. Rev.} D}

\def\ra{\rightarrow}

\def\nuebar{\bar\nu_e}
\def\numubar{\bar\nu_\mu}
\def\nue{\nu_e}
\def\anue{\bar\nu_e}
\def\nutau{\nu_\tau}
\def\anutau{\bar\nu_\tau}
\def\anumu{\bar\nu_\mu}
\def\numu{\nu_\mu}

\def\anul{\bar\nu_\ell}
\def\nul{\nu_\ell}

\begin{document}

\thispagestyle{empty}
\begin{flushright}
{\tt DFPD 00/EP/49}\\ 
{\tt ICARUS/TM-2000/04}\\ 
\today
\end{flushright}
\vspace*{1cm}
\begin{center}
{\Large{\bf The flavor of neutrinos
in muon decays
\\at a neutrino factory and the LSND puzzle} }\\
\vspace{.5cm}
A. Bueno$^{a,}$\footnote{Antonio.Bueno@cern.ch},
M. Campanelli$^{a,}$\footnote{Mario.Campanelli@cern.ch},
M. Laveder$^{b,}$\footnote{Marco.Laveder@pd.infn.it},
J. Rico$^{a,}$\footnote{Javier.Rico@cern.ch},
A. Rubbia$^{a,}$\footnote{Andre.Rubbia@cern.ch}

\vspace*{0.3cm}
$~^a$Institut f\"{u}r Teilchenphysik, ETHZ, CH-8093 Z\"{u}rich,
Switzerland\\
$~^b$ Dip. di Fisica ``G. Galilei'', Univ. di Padova and INFN Sez. di Padova, 
I-35131 Padova, Italy

\end{center}
\vspace{0.5cm}

\setcounter{footnote}{0}

\abstract{The accurate prediction of the neutrino beam produced in
muon decays and the absence of opposite helicity contamination for
a particular neutrino flavor
make a future neutrino factory the ideal place to look for 
the lepton flavor violating (LFV) decays of the kind 
$\mu^+\ra e^+\nuebar\numu$
and lepton number violating (LNV) processes like $\mu^-\ra e^-\nue\numu$.
Excellent sensitivities can be achieved using a detector
capable of muon and/or electron identification with charge 
discrimination. This would allow to set experimental limits that 
improve current ones by more than two orders of magnitude and test 
the hypothesis that the LSND excess is due to such anomalous 
decays, rather than neutrino flavor oscillations in vacuum.}
\newpage

\section{Introduction}

The Standard electroweak Model (SM) is built on
an absolute conservation of three separate
lepton flavors (LF): the electron flavor $L_e$, the muon flavor $L_\mu$,
and the tau flavor $L_\tau$. The total lepton number is defined as 
$L_{tot} = L_e + L_\mu + L_\tau$.

In a model independent way, the main decay mode of the positive
muon\footnote{Similar arguments hold for negative muons.} into
a positron and two neutrinos can be written
as\cite{Herczeg:1992qy,Herczeg:1997bu}:
\begin{equation}\label{eq:mudec}
\mu^+\ra e^+ +n+n'
\end{equation}
where $n$ and $n'$ denote neutrinos, which can be either
neutrinos or antineutrinos and of any flavor $e$, $\mu$
or $\tau$.
In the SM, individual LF conservation implies that $n\equiv\nue$ and
$n'\equiv\anumu$.

The evidence from neutrino oscillation searches
that the neutrinos are in fact massive and mixed,
implies that the LF conservation is not
exact. 
However, LF conservation is constrained by
stringent experimental limits obtained in processes
involving charged leptons. 
For example, the present experimental 90\%C.L. upper limits 
on the most
interesting of these decays are:
\begin{eqnarray}
BR(\mu \ra e \gamma) \, < \, 1.2 \times 10^{-11} \, &:& \, \,
\cite{Brooks} \\
BR(\mu^+ \ra e^+ e^+ e^-) \, < \, 1.0 \times 10^{-12} \, &:& \, \,
\cite{Bellgardt} \\
R(\mu^- Ti \ra e^- Ti) \, < \, 6.1 \times 10^{-13} \, &:& \, \,
\cite{Wintz} \\
BR(\tau \ra \mu \gamma) \, < 1.1 \times \, 10^{-6} \, &:& \, \,
\cite{CLEO}
\end{eqnarray}
These stringent limits are not inconsistent with the neutrino
oscillation results since given reasonable upper bounds on the
neutrino masses, the effect induced within the SM would
be too small to be seen\cite{Herczeg:1998si}. However, within
extensions of the SM, neutrino oscillations raised the
possible prospect that there might exist observable processes
that violate the charged-lepton number\cite{Ellis:2000uq}.
Projects are currently underway to improve several of these upper limits
significantly. 

In contrast to this, direct experimental 
limits on LF conservation in decays involving
neutrinos are much less
stringent. For example, the limit
\begin{equation}
BR(\mu^-\ra e^-\nue\numubar) < 1.2\times 10^{-2} \, : \, \,
\cite{Freedman:1993kz}
\end{equation}
is orders of magnitude worse than limits involving
charged leptons. 
In extensions of the SM, the existence of LF violation
could open new modes of decay where
$n$ and $n'$ in Eq.~(\ref{eq:mudec}) are indeed
any given neutrinos or antineutrinos.

One would clearly like
to experimentally test these limits 
by a sensitive study of the neutrinos
produced in such decays. 
Such studies could directly be relevant to a further understanding
of the neutrino sector, as
described in the following sections and would represent
complementary investigations to those for neutrino
flavor oscillations.

If the excess of events found in the Los Alamos LSND experiment\cite{DAR,Mills} 
is interpreted as 
due to $\mu^+\ra e^+ +\nuebar+n$ (with a branching ratio equal to the 
measured probability of $\anue$ appearance in case of flavor neutrino 
oscillations), this anomalous decay could be easily tested at a future 
neutrino factory with a detector capable of charge discrimination,
looking at the more convenient decay $\mu^-\ra e^-\nue\numu$ as
described in section~\ref{sec:lvn}. 

Good sensitivities could also be reached at a neutrino factory for the
search $\mu^+\ra e^+\anul\numu$ described in section~\ref{sec:lfv} 
in which one look for
interactions of $\numu$ giving $\mu^-$ in a pure beam that produces $\mu^+$.

\section{The neutrino oscillation sector}
Hints that neutrinos are massive particles come from the observation
of three 
anomalous effects --- the LSND excess\cite{DAR,DIF,Mills}, 
the atmospheric anomaly\cite{Kam-atm,IMB,SK-atm,Soudan2,MACRO} and the solar
neutrino
deficit\cite{Homestake-98,Kamiokande-sun-96,GALLEX-99,SAGE-99,SK-sun}.
In particular, the atmospheric results are the most convincing ones. 
All three effects can be naturally explained
in terms of neutrino flavor oscillations, which
will occur when neutrinos propagate through space, if their
masses are non-degenerate and the weak and mass eigenstates are 
mixed\cite{pontecorvo}. 

However, in order to explain {\it all} three experimentally 
observed effects in terms of neutrino flavor 
oscillations, one is forced
to invoke additional sterile neutrino states\cite{Giunti:2000bn} 
to accommodate the very different frequencies of oscillations
--- given by the mass differences squared $\Delta
m^2$'s --- indicated by the three different effects.
The existence of such neutrinos is a currently unresolved problem
and clearly demonstrates that the neutrino sector is not
fully understood.
Several attempts were made to explain all data in terms
of only three massive neutrinos\cite{threenus}. But they are all
excluded by the latest data.

From a phenomenological point of view, we recall that the neutrino flavor 
oscillation hypothesis predicts
a well defined dependence of the phenomenon as a function of the neutrino
energy, characterized by the so-called $L/E$ behavior, where
$L$ is the distance between the source and detector and $E$ the neutrino
energy. So far, no experiment has conclusively demonstrated
such a $L/E$ dependence of the anomalous effect, with maybe 
the exception of the SuperKamiokande data
which favors\cite{Sobel_00} a dependence $\propto LE^{n}$ where $n\approx -1$.

In such an unclear situation, is it possible to 
envisage ``non-flavor-oscillation''
mechanisms to explain part of the neutrino data?

Aside from theoretical arguments against sterile neutrinos, 
we argue that, from a phenomenological point of view,
the LSND effect is particular: it has a small
probability, measured to be $(2.5 \pm 0.6 \pm 0.4)\times 10^{-3}$\cite{Mills},
in contrast to the solar and atmospheric
neutrino anomalies, which are large. Hence, LSND is a
natural candidate for an interpretation involving a
different physics than in
atmospheric and solar neutrino flavor oscillations.

\section{Relevance to the LSND puzzle}

We recall that the LSND effect was first reported as an excess
of $\anue$'s in the $\anumu$'s flux from the $\mu^+$ Decay-At-Rest (DAR) 
process\cite{DAR}. The neutrino beam is obtained with 800~MeV kinetic energy protons 
hitting a series of targets, producing secondary pions. 
Most of the $\pi^+$ come to rest
and decay through the sequence $\pi^+\ra\mu^+\numu$, followed
by $\mu^+\ra e^+\nue\numubar$, supplying the experiment with
the $\numubar$'s with a maximum energy of 52.8~MeV. The intrinsic
contamination of $\bar\nue$'s coming from the symmetrical
decay chain starting with $\pi^-$ is estimated to be small since
most negatively charged mesons are captured before they decay.

The excess of $\anue$'s, explained in terms of neutrino
flavor transitions of the type $\numubar\ra\nuebar$,
occurs via the reactions:
\begin{eqnarray}
 \mu^+\ra e^+\nue\numubar;\ \ \ \ \ \ 
 \anumu\stackrel{vacuum}{\longrightarrow}\anue;\ \ \ \ \  \anue p \ra e^+ n
\end{eqnarray}
Additional evidence in favor of neutrino
oscillation was reported in the Decay-In-Flight (DIF) sample,
though with a low statistical significance\cite{DIF}. In a recent re-analysis
of the complete data sample collected\cite{Mills}, the 
significance of the DIF data seems to be even lower. Hence, we concentrate
on the hint from stopped muons, and ignore the DIF result.

The latest KARMEN2 results\cite{Eitel:2000by} come very close to contradicting the
LSND claim, however the experimental sensitivity is marginal to 
conclusively exclude or confirm completely the LSND excess.
A new experiment, MiniBOONE\cite{Bazarko:2000id},
will confront the flavor oscillation hypothesis with a very
high statistical accuracy. 

In case of a negative result, one will
only be able to conclude that the LSND excess was not due to neutrino flavor
oscillations.

The implications of exotic muon decays on the LSND excess
has been studied\cite{Herczeg:1997xi} showing
that two explicit models predicted interactions of about
one order of magnitude smaller than what would be
relevant for LSND.
In a model independent proof in which any contributions from 
neutrino mixing is neglected, the authors of Ref.\cite{Bergmann:1999ft} 
prove that new lepton flavor violating 
interactions, under the constraint of LF data involving
charged leptons, fail short to explain the LSND effect by
a rate factor of almost three. However in Ref.\cite{Bergmann:2000gn}, 
it is reported that exotic decays that produce two antineutrinos 
\begin{equation}
\mu^+\ra e^+ + \anue + \anul\ \ \ \ \ (\ell=e,\mu,\tau)
\end{equation}
cannot be ruled out as the cause of the LSND excess in
a model independent way.

Regardless of any theoretical prejudice, 
the excess of electrons found by the LSND 
experiment stands today as a still unresolved puzzle
of neutrino physics.

\section{Searches at a neutrino factory}
A neutrino factory\cite{geers,nufacwww} is understood as a machine where
low energy muons of a given charge are accelerated in a storage
ring. The two neutrinos $n$, $n'$ produced in the decay Eq.~(\ref{eq:mudec})
will be boosted
in the forward direction of the muon flight path. Hence, the
muon storage ring is composed of long straight-lines in order
to produce directional neutrino beams. 

In such machines, muons are produced in decays of secondary pions
produced by few GeV protons incident on a target. In current designs, muons
are captured with high efficiency and very high integrated
protons-on-target intensities are envisaged in order to produce 
very intense neutrino
sources.

The neutrino physics potentialities of such machines has been largely discussed
in the literature\cite{Albright:2000xi,nufacfis}. 
In particular, we mention our study in the context of a short-baseline 
experiment to search for neutrino flavor oscillation in a background free 
environment in the $\Delta m^2$ region indicated by LSND\cite{Bueno:1998xy}. 

A neutrino factory is also an ideal place to study neutrinos
from exotic muon decays.
The envisaged flux of neutrinos is sufficiently high to obtain
large statistics of neutrino interaction events. More importantly, 
we can take advantage from the fact that the neutrino beam
is produced from muons of a definite sign and therefore the decay
processes can be studied with {\it a pure initial state}. This is not
the case for traditional pion decay neutrino beams, in which contaminations
are always present at some level.

The flavor of the interacting neutrinos can be tested via their
charged current  processes.
In case of purely lepton flavor conserving decays
$\mu^+\ra e^+\nue\numubar$, we expect to detect only
\begin{eqnarray}
\nue + N \ra e^- + X \\
\numubar + N \ra \mu^+ + X
\end{eqnarray}
while exotic decays can be immediately identified by various 
processes
\begin{eqnarray}
& \mu^+\ra e^+ + n + \numu \longrightarrow\ \ \ \ \ \ 
& \ \ \ \ \  \numu + N \ra \mu^- + X \\
& \mu^+\ra e^+ + \anue+n \longrightarrow\ \ \ \ \ \ 
& \ \ \ \ \  \anue + N \ra e^+ + X \\
& \mu^+\ra e^+ + n + \nutau \longrightarrow\ \ \ \ \ \ 
& \ \ \ \ \  \nutau + N \ra \tau^- + X \\
& \mu^+\ra e^+ + n+\anutau \longrightarrow\ \ \ \ \ \ 
& \ \ \ \ \  \anutau + N \ra \tau^+ + X 
\end{eqnarray}
where $n$ stands for neutrinos or anti-neutrinos. 

Charge discrimination of electrons and muons can trivially separate the
two types of decays. The presence of taus is more difficult
to identify but can be achieved using topological or
kinematical signatures. It requires neutrino beams of
high energy in order to exceed the tau lepton production threshold.
We do not consider tau identification any further in this
paper, and concentrate on the identification of electrons
and muons in a low energy setup.

In order to predict the energy distribution of the (anti)neutrinos in
the detector we assume for definiteness two types of generic decays: 
\begin{eqnarray}
& \mu^+\ra e^+ + \anul + \numu \label{eq:dk1}\\
& \mu^-\ra e^- + \nue + \nul \label{eq:dk2}
\end{eqnarray}

We address the search for LFV, Eq.~(\ref{eq:dk1}), 
using wrong sign muons, which are
experimentally simpler to detect. In order to profit from the enhanced
cross-section of neutrinos versus antineutrinos, it is better to
select positive muons in the storage ring, since in this case, the LFV
decays produce $\numu$'s.

In case we consider the LNV decay of Eq.~(\ref{eq:dk2}), for which 
$\Delta L=2$, we should look for wrong sign electrons. We select 
$\mu^-$ in the ring since the signal searched for in this case has two 
neutrinos in the final state, therefore we profit from the enhanced
neutrino cross sections. Naturally, both signs of muons could be
studied in a real experiment, in order to provide possible checks
for different behaviors in $\mu^+$ or $\mu^-$ decays.

Following the discussion in~\cite{Bergmann:2000gn}, LNV interactions 
can naturally arise via mixing of 
heavy bosons that transform differently under the Standard Model group 
but identically under the unbroken $U(1)_{EM}$. In particular, 
the effective four-fermion operator relevant for the reaction: 
\begin{equation}
\mu^-_L\ra e^-_R + \nue + \nul
\label{eq:lnv}
\end{equation}
which violates lepton 
number conservation by two units, has the form 
($\mu_L\bar{\nu}_e$)($\bar{\nu}_\ell\bar{e}_R$). It couples $\nu_e$ to 
$\mu^-$ and $e^-$ to $\nul$. This operator can be induced, for example in 
supersymmetric models without R-parity, through the mixing of sfermions 
that are $SU(2)_L$ singlets with sfermions that are $SU(2)_L$ doublets. 

\begin{figure}[p]
\centering
\epsfig{file=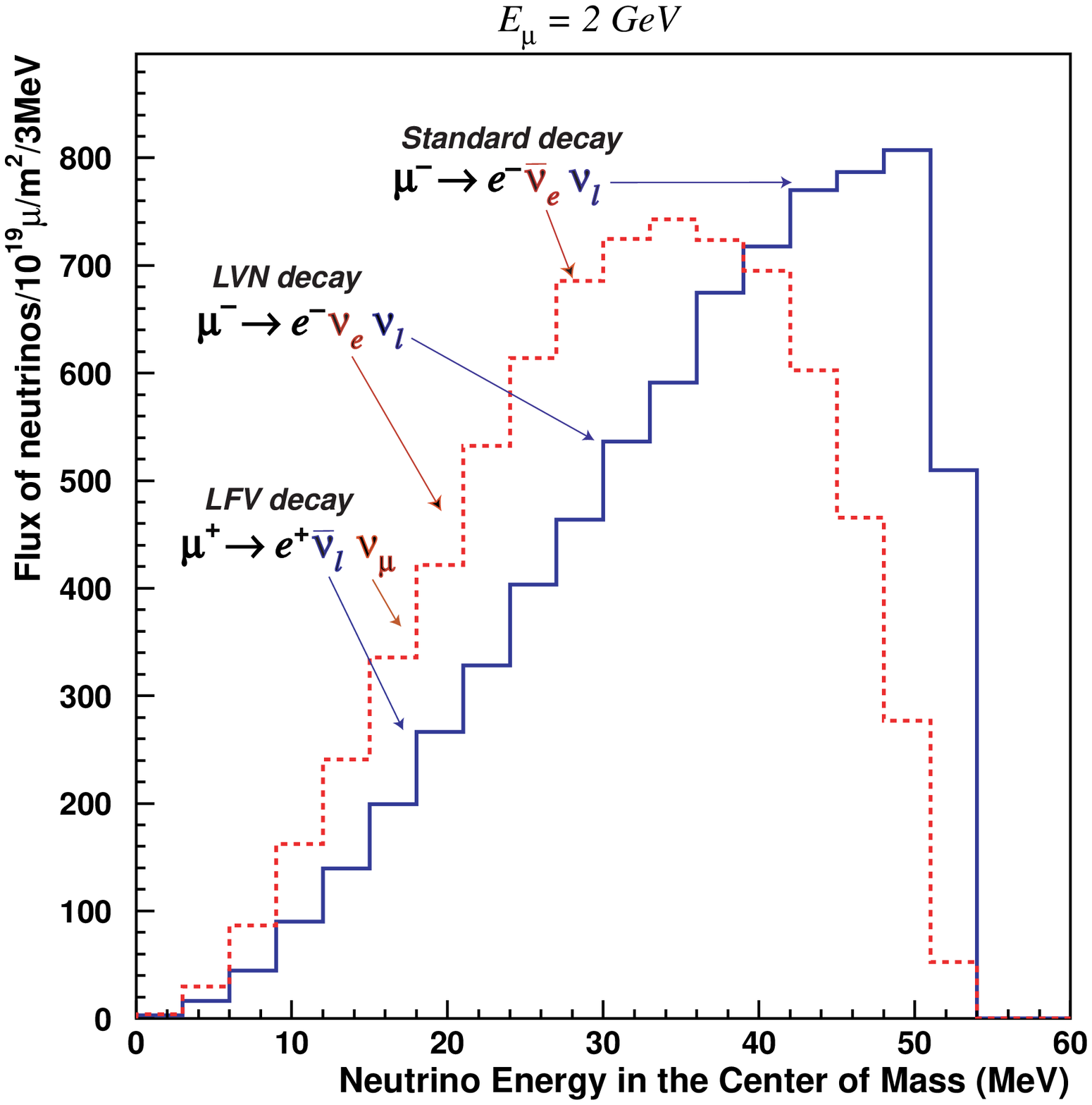,width=14cm}
\caption{Expected neutrino fluxes for the standard, lepton flavor and 
lepton number violating $\mu^-$ decays as a function of the neutrino 
energy in the center of mass reference frame.}
\label{fig:lnvspectra}
\end{figure}

After explicit calculation, we obtain that the square of the
scattering amplitude for the decay~(\ref{eq:lnv}) is 
$<|{\cal M}|^2>\;\propto\; (p_\mu \cdot p_{\nue})(p_e \cdot
p_{\numu})$ (where $p_i$ is the four-momentum of
particle $i$). This expression is similar to the one obtained for 
the standard decay $\mu^-\to e^-\bar{\nu}_e\nu_\mu$; therefore the flux 
of $\nu_e$'s coming from (\ref{eq:lnv}) is equal to the one of 
$\bar{\nu}_e$'s produced in the standard $\mu^-$ decay as shown in 
Figure~\ref{fig:lnvspectra}.

\section{Experimental considerations}
We address a few experimental considerations in the context
of an optimization of searches for LFV and LNV decays of the muon.

\subsection{Beam setup}
Unlike for neutrino flavor oscillations in vacuum, the distance $L$
between source and detector is in this case 
an irrelevant physical parameter. It is hence advantageous
to place the experiment close to the source in order to
gain flux like $1/L^2$ due to the beam divergence.

We think that a neutrino beam energy of a few GeV would be optimal
in order to facilitate the discrimination of the muon and electron
charges, and in order to reduce misidentified electrons or muons
background coming mostly from neutral current interactions.
We therefore consider a low-energy muon storage ring\footnote{We
also note that studies indicate that the
cost of a neutrino factory is driven by the muon
energy $E_\mu$ and rises very rapidly with $E_\mu$, hence
a low energy muon beam is also financially favored.} with three possible
muon energies $E_\mu=1$, $2$ or $5\rm\ GeV$.

\begin{figure}[p]
\centering
\epsfig{file=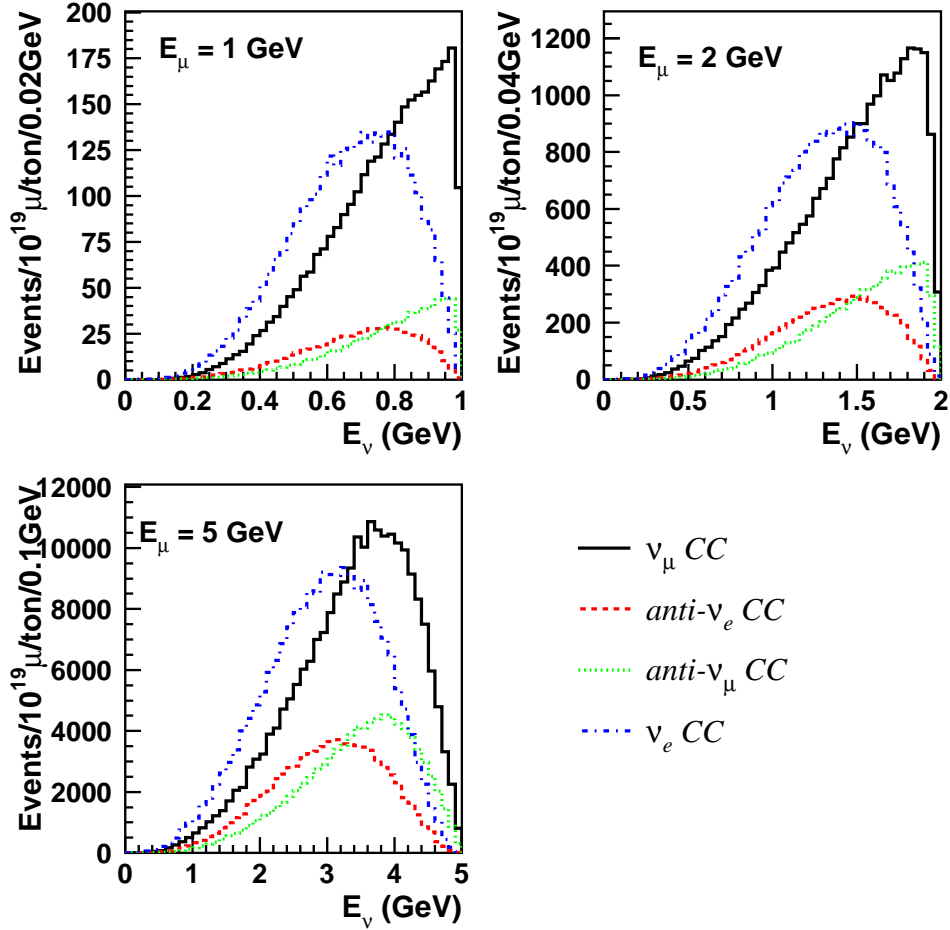,width=14cm}
\caption{Neutrino events energy spectrum for muon ring energy of 1, 2 and
5~GeV. The $\numu$ (line) and $\anue$ (dashed) events come from standard $\mu^-$ decays,
while $\anumu$ (dotted) and $\nue$ (dash-dotted) events come from standard $\mu^+$ decays.}
\label{fig:evrate}
\end{figure}

In Table~\ref{tab:rates}, we list the expected event rates from standard
muon decays per ton of target and $10^{19}$ standard muon decays. The detector
is located at a distance $L=100\rm\ m$ from a 100~m long straight section
of the storage ring. The expected event energy spectra are shown
in Figure~\ref{fig:evrate}.

\begin{table}[tb]
\begin{center}
\begin{tabular}{|cc|c|c|c|}
\hline\hline
 & & $E_\mu=1$ GeV & $E_\mu=2$ GeV & $E_\mu=5$ GeV \\
\cline{3-5}
 & $\numu$ CC                 & 3300 & 23200 & 233400 \\
$\mu^-$ & $\numu$ NC          & 410 & 4470   & 58600 \\
$10^{19}$ decays & $\anue$ CC & 630 & 6250   & 80200 \\
 & $\anue$ NC                 &  100 & 1450  & 23100  \\ \hline\hline
 & $\anumu$ CC                & 730 & 7200   & 91200 \\
$\mu^+$ & $\anumu$ NC         & 140 & 1830   & 27600 \\
$10^{19}$ decays & $\nue$ CC  & 3060 & 21500 & 211700 \\
 & $\nue$ NC                  & 310 & 3660   & 50100 \\ \hline
\end{tabular}
\caption{Expected event rates (CC=charged current, NC=neutral currents)
per ton of target per $10^{19}$ standard muon decays
in a storage ring with a straight line of 100~m and
located 100~m away from the neutrino detector. Only 50\% of the muons are assumed to
decay in the direction of the detector, the other 50\% are lost.}
\label{tab:rates}
\end{center}
\end{table}

\subsection{Detector parameters}
Experimentally, the presence of LFV decays will characterize themselves
by the observation of events with ``wrong sign leptons''.
It is therefore mandatory to ensure a very good and efficient determination
of the lepton charge.

The muon charge is most easily determined with the help of bending in a 
magnetic field. 
The radius of curvature in meters
in a 1~T field is approximately $10p(GeV)/3$, 
or 3~meters for $p=1\rm\ GeV$. In the case of electrons, the capability
to measure the charge is limited by the radiation length of the
target which determines the distance after which the electromagnetic
shower develops at a level where the primary electron is not distinguishable
any longer.

At low energies, we expect events to exhibit simple
topologies and many of them will be quasi-elastic-like.
Hence, the events will be dominated by a leading hard lepton 
accompanied by few soft
hadrons. Such events were traditionally best studied in bubble chambers,
due to the required low density, high granularity and homogeneity to
capture soft escaping tracks at all angles. In addition, a low density,
high granularity instrumented target 
is mandatory to efficiently recognize electrons and
to discriminate muons from pions. Different detector configurations
meeting the previously mentioned specifications can be clearly envisaged.

\section{Results for $\mu^+\ra e^+ + \anul + \numu$}
\label{sec:lfv}

We consider a 10 ton fiducial mass detector located at a distance of
100~m from the muon storage ring. 

For a 2~GeV muon ring energy, the expected event samples for a total
of $10^{19}$ standard $\mu^+$ decays are 72'000 $\anumu$ CC and 
215'000 $\nue$ CC events, and 18'300 $\anumu$ NC and 36'600 $\nue$ NC.

\begin{table}[tb]
\begin{center}
\begin{tabular}{|c|c|c|c|c|}
\hline
Cuts & $\anumu$ CC & $\anumu$ NC & $\nue$ NC & LFV $\numu$ CC \\
\hline\hline
 \multicolumn{5}{|c|}{$E_{\mu^+}=2\rm\ GeV$}\\
\hline
Initial & 72000 & 18300 & 36600 & 540 \\
$\mu^-$ candidate & 9890 & 660 & 1197 & 540 \\
$E_{\mu^-}>1.1$ GeV & $<0.1$ & 0.6 & 2 & 130 \\ 
\hline\hline
 \multicolumn{5}{|c|}{$E_{\mu^+}=5\rm\ GeV$}\\
\hline
Initial & 912000 & 276000 & 501000 & 5290 \\
$\mu^-$ candidate & 133500 & 10500 & 17900 & 5290 \\
$E_{\mu^-}>3$ GeV & 3 & 0.3 & 0.6 & 645 \\ 
\hline\hline
 \multicolumn{5}{|c|}{$E_{\mu^+}=1\rm\ GeV$}\\
\hline
Initial & 7300 & 1400 & 3100 & 76 \\
$\mu^-$ candidate & 873 & 46 & 99 & 76 \\
$E_{\mu^-}>0.6$ GeV & $<0.1$ & 0.1 & 0.6 & 16 \\ \hline
\hline
\end{tabular}
\caption{Effect of cuts on background and signal. We assumed
a positive muon ring energy $E_{\mu^+}$ of 
1, 2 and 5~GeV and a total of $10^{19}$ standard
decays. The LFV decay has the branching probability of $2.5\times 10^{-3}$.
Backgrounds come from hadrons escaping the detector without interacting
or muons from meson decays. For the charged current background, no veto
on the positive muon has been included. $E_{\mu^-}$ is the energy
of the identified negative muon in the event.}
\label{tab:analysis}
\end{center}
\end{table}

To estimate the signal efficiency, we assume that the LFV decay proceeds
through a similar diagram as the standard muon decay as given by the $V-A$
theory, however, with interchanged neutrino flavors. We can then
essentially assume that {\it at the detector location} 
the flux of $\numu$'s from LFV decays 
is similar
to the one of $\nue$'s in the standard muon decay. It is clear that other
type of interactions could be envisaged and could lead to
different energy spectra which can be experimentally tested by studying
the visible energy distribution of wrong sign muon events.

If LFV decay occur with a branching $Br(LFV)$, we expect to observe
the number of negative muon events
\begin{eqnarray}
N_{\mu^-,CC} & = & \int \Phi_{LFV}(\numu,E_\nu) \sigma^{CC}(\numu,E_\nu)dE_\nu \\
& \approx & Br(LFV)\times \int \Phi(\nue,E_\nu)
\sigma^{CC}(\numu,E_\nu)dE_\nu 
\end{eqnarray}
where $\Phi_{LFV}$ is the flux of $\numu$ neutrinos from LFV decays,
$\Phi(\nue)$ is the flux of electron neutrinos in standard $\mu^+$
decays and $\sigma^{CC}$ the charged current cross-sections.

For the branching indicated by LSND and $10^{19}$ $\mu^+$ decays, we obtain
\begin{eqnarray}
N_{\mu^-,CC} & \approx & 215'000 \times Br(LFV) \\
& = & 540
\end{eqnarray}
for the fiducial detector mass of 10~ton.

For muon tracks of a few GeV momentum bent in the magnetic field, we
expect wrong charge confusion at the level of $\approx 10^{-3}\%$. We therefore
think that the background produced by $\bar\numu$ CC events with
mismeasured charge of the muon to be less than one event.

One crucial experimental aspect is the discrimination of muons versus
pions. In addition, secondary muons from meson decays produce
background, especially at low energies.
In order to assess realistic efficiencies and experimental backgrounds,
we illustrate results with a detector with the characteristics
of the ICARUS
liquid Argon imaging TPC\cite{Vignoli:2000yn}. The detector would be
on a medium-sized $LAr$ vessel surrounded by a dipole magnet.
The size of such detector would be in the range of $2\times 2\times 6\rm\
m^3$ for a total mass of about 30 tons. 
The magnetic field would be oriented perpendicular to the
drift $E$-field and the incoming neutrino direction, in order
to bend the charged particles 
in the direction of the drift
field, where a resolution in the range of $200\rm\ \mu m$ is expected
from the LAr TPC.
Given the hadronic interaction length of $LAr$ of $\lambda_I = 84\rm\ cm$,
muons which loose about $240\rm\ MeV/m$ are distinguished
from pions which interact hadronically. 

The effect of simple cuts on background and signal are illustrated
in Table~\ref{tab:analysis}. We assumed in this case
a positive muon ring energy and a total of $10^{19}$ standard
decays. The LFV decay has the branching probability of $2.5\times 10^{-3}$,
compatible with the LSND excess.

We considered all sources of backgrounds by means of fully generated
neutrino events. Large event samples were produced for all neutrino 
species with the proper energy distribution with the help
of an event generator\cite{nux} which includes all exclusive final states
and a realistic treatment of the low energy region. 

A muon candidate is identified as a track which stops without
interacting in the Argon or leaves the detector vessel before
interacting or stopping.
Hence, charged and neutral current backgrounds come from hadrons 
escaping the detector without interacting or
or from actual muons from meson decays. 
For the charged current background, no veto
on the positive muon has been included.

\begin{figure}[p]
\centering
\epsfig{file=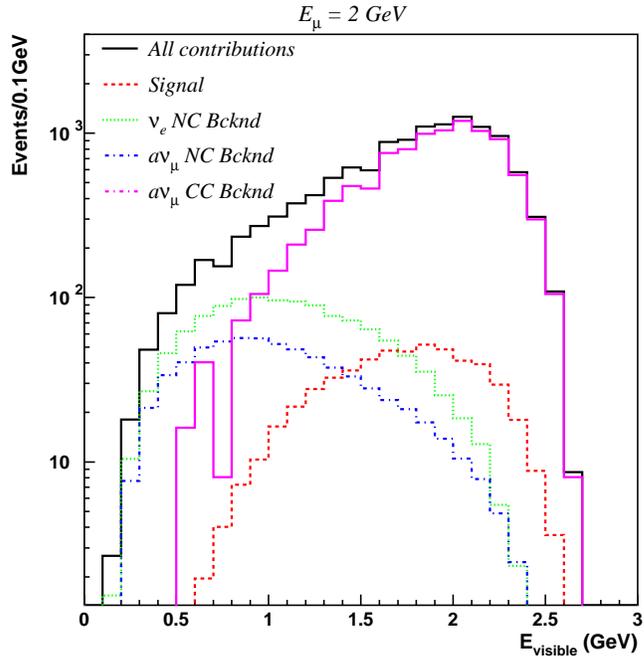,width=9.5cm}\\
\epsfig{file=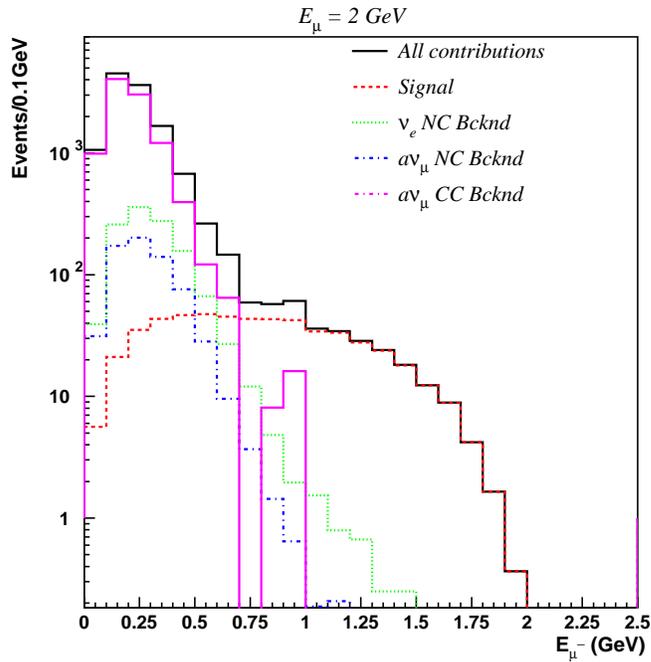,width=9.5cm}
\caption{Visible energy  (upper) and
candidate muon (lower) distribution for LFV decays (see text)
normalized to LSND excess and $10^{19}$ positive
muon decays for 10 ton detector. The background
from neutral current using the characteristics of an
ICARUS $LAr$ TPC is also shown.}
\label{fig:distrnocut}
\end{figure}

\begin{figure}[p]
\centering
\epsfig{file=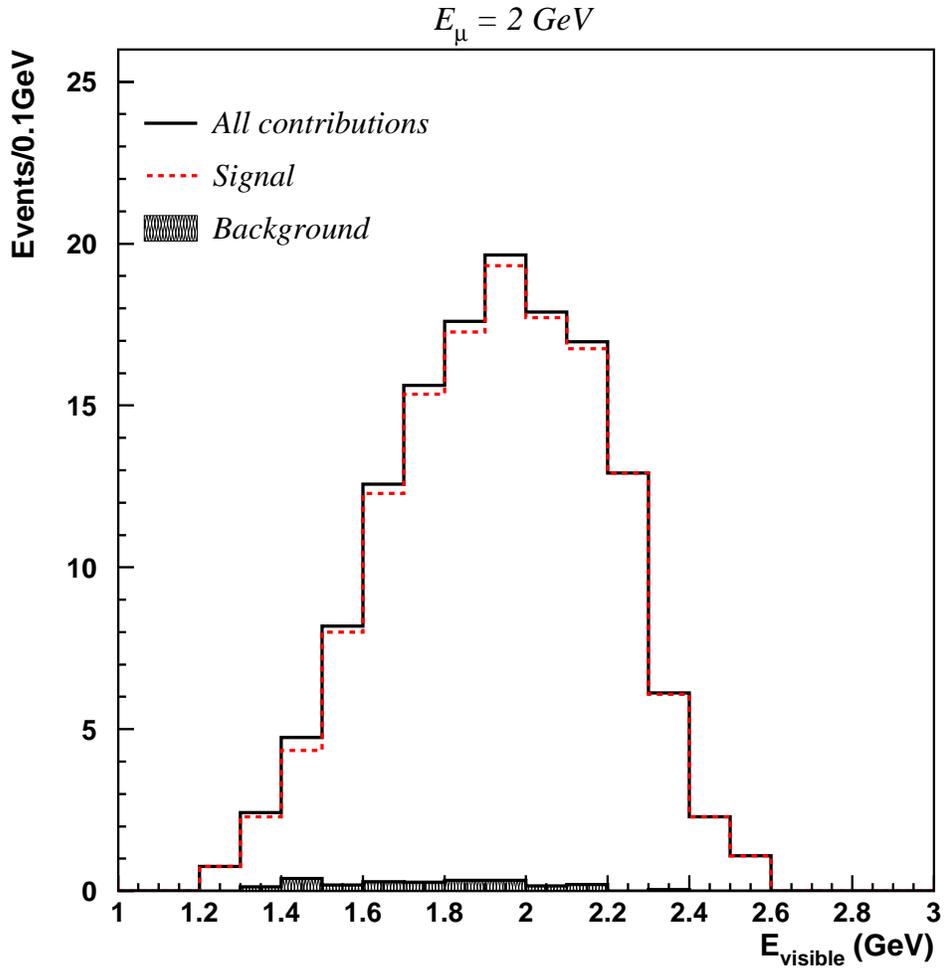,width=14cm}
\caption{Visible energy distribution (see Figure~\ref{fig:distrnocut})
 after a cut on the candidate muon momentum for LFV signal (see text)
and backgrounds expected in a detector
with the characteristics of an ICARUS $LAr$ TPC.}
\label{fig:distrcut}
\end{figure}

In Table~\ref{tab:analysis}, we observe that the presence
of LFV decays will produce an excess of events with
$\mu^-$ candidates. The visible energy and the candidate muon
energy spectra are shown in Figure~\ref{fig:distrnocut}. As
expected, the excess is most visible at the highest end
of the muon energy distribution, since background comes
from misidentified soft hadrons. A simple cut on the
muon candidate energy can be used to largely suppress
background to negligible level, while keeping a large
fraction of the LFV signal events, as shown in the
Table~\ref{tab:analysis}. The visible energy distribution
of events for $E_{\mu^+}=2\rm\ GeV$ with a cut on the
muon candidate momentum of 1.1~GeV (efficiency for LFV
of $25\%$) is shown in Figure~\ref{fig:distrcut}. The
statistics is clearly sufficient to constrain any
theoretical prediction of the energy distribution
of the LFV neutrinos.

\begin{table}[ht]
\begin{center}
\begin{tabular}{|c|c|c|}
\hline
Cuts & $\nue$ CC & $\anue$ CC \\
\hline\hline
 \multicolumn{3}{|c|}{$E_{\mu^-}=2\rm\ GeV$}\\
\hline
Initial & 540 & 62500 \\
One proton & 367 & 11000 \\
No pions & 323 & 100 \\
$E_e >1$ GeV & 103 & 17 \\ 
Candidate charge & 21 & 0.4 \\
\hline\hline
 \multicolumn{3}{|c|}{$E_{\mu^-}=5\rm\ GeV$}\\
\hline
Initial & 5290 & 802000 \\
One proton & 3390 & 212160 \\
No pions & 2112 & 495 \\
$E_e >3$ GeV & 351 & 163 \\ 
Candidate charge & 71 & 4 \\
\hline\hline
 \multicolumn{3}{|c|}{$E_{\mu^-}=1\rm\ GeV$}\\
\hline
Initial & 76 & 6300 \\
One proton & 53 & 529 \\
No pions & 48 & 8 \\
$E_e >0.2$ GeV & 43 & 4 \\ 
Candidate charge & 10 & 0.1 \\
\hline\hline
\end{tabular}
\caption{Effect of cuts on background and signal. We assumed
a negative muon ring energy $E_{\mu^-}$ of 
1, 2 and 5~GeV and a total of $10^{19}$ standard
decays. The lepton number violating decay has the branching 
probability of $2.5\times 10^{-3}$. 
$E_e$ is the energy of the identified electron in the event.}
\label{tab:elesearch}
\end{center}
\end{table}

\section{Results for $\mu^-\ra e^- + \nue + \nul$}
\label{sec:lvn}

The experimental signal consists in the appearance of final state 
electrons, while standard events have positrons in the final state. 
Therefore, to experimentally detect reaction (\ref{eq:dk2}), it is mandatory 
to envisage an experiment endowed with charge discrimination 
capabilities for electrons. This is a true experimental challenge given 
the short radiation lengths in dense targets.

In order to evaluate realistic efficiencies and backgrounds, we consider 
an experimental setup similar to the one discussed in the previous section, 
namely, an ICARUS liquid Argon imaging TPC with a magnetic field provided 
by an external dipole magnet.

Table~\ref{tab:elesearch} shows the effect of the cuts applied
for a normalization of $10^{19}$ muon decays. 
To compute the expected number of signal events  we
have taken a branching probability of $2.5\times 10^{-3}$.
Given the low muon energies considered, most of the events will be 
quasielastic. We thus require a final state configuration containing 
an electron and a reconstructed proton\footnote{We assume a kinetic
threshold of $T>50\rm\ MeV$ in order to detect a proton.} 
and no additional hadronic particles. 
These criteria reduce the quasielastic background (where we expect
a neutron rather than a proton in the final state)
by almost three orders of magnitude,
while keeping more than $50\%$ of the signal. The NC background, where 
electron candidates come from $\pi^0$ conversion, is in general soft. After a 
cut on the electron candidate momentum, this kind of background becomes 
negligible.

As mentioned before, charge discrimination between electrons and positrons 
is a must for this kind of search. Preliminary studies show\cite{elefit} 
that for a LAr detector immerse in 
a 1T magnetic field, a fit to the direction of the electromagnetic shower 
could provide a good determination of the charge of leading electrons. 
When loose criteria are applied, the expected charge contamination from 
$\anue$ CC amounts to $2\%$ for an electron efficiency of 25$\%$. Tighter
requirements reduce the charge confusion to the per mil level for an electron 
identification efficiency close to $10\%$. Table~\ref{tab:elesearch} shows 
that applying loose criteria in the determination of the lepton charge is 
enough to eliminate $\anue$ CC background.

\section{Experimental sensitivities}
\label{sec:sensi}

Table~\ref{tab:limits} shows the expected sensitivities in case a negative
result is found as a function of the muon energy circulating in 
the accumulation ring. Three different normalizations have been 
considered: $10^{18}$,$10^{19}$ and $10^{20}$ muons. For comparison 
we also show the current best limit on the LFV decay 
$\mu^+ \to e^+\nu_e\bar{\nu}_\mu$~\cite{Freedman:1993kz}.

For a statistics corresponding to $10^{19}$ muon decays we 
could improve the present 
sensitivity by more than two orders of magnitude. Three orders of magnitude 
can be reached for $10^{20}$ muons. 
Therefore the interpretation of the LSND excess in terms of 
anomalous muon decay that violates lepton flavor and/or total lepton
number could be thoroughly experimentally tested.

\begin{table}[htb]
\begin{center}
\begin{tabular}{cccccc}\hline\hline
$\mu$ Decays & Decay mode & Current Limit & $E_\mu = 1$ GeV & $E_\mu = 2$ GeV & $E_\mu = 5$ GeV \\ \cline{4-6}
$10^{18}$ & & & $4\times 10^{-3}$ & $5\times 10^{-4}$ & $1\times 10^{-4}$ \\

$10^{19}$ & $P(\mu^+ \to e^+\anul\nu_\mu)<$ & $1.2\times 10^{-2}$ & $5\times 10^{-4}$ & $1\times
10^{-4}$ & $3\times 10^{-5}$ \\

$10^{20}$ & & & $2\times 10^{-4}$ & $6\times
10^{-5}$ & $2\times 10^{-5}$ \\ \hline

$10^{18}$ &  &  & $6\times 10^{-3}$ & $3\times
10^{-3}$ & $9\times 10^{-4}$ \\ 

$10^{19}$ & $P(\mu^- \to e^-\nu_e\nul)<$ & $-$ & $6\times 10^{-4}$ & $3\times
10^{-4}$ & $3\times 10^{-4}$ \\ 

$10^{20}$ &  &  & $1\times 10^{-4}$ & $1\times
10^{-4}$ & $2\times 10^{-4}$ \\\hline\hline
\end{tabular}
\caption{Achievable limits in case of negative result
at the $90\%C.L.$ for $\mu^+\ra e^+\anul\numu$ and $\mu^-\ra
e^-\nue\nul$ decays
with a 10~ton detector for three different number of muon decays.}
\label{tab:limits}
\end{center}
\end{table}

\section{Conclusion}
A negative result from the MiniBOONE experiment would
indicate that the neutrino flavor oscillation is not the correct
hypothesis to explain the excess seen in LSND. It would however
not contradict other possible non-flavor-oscillation interpretations
of the effect.

In particular, LFV and LVN decays could play a role in the interpretation
of the LSND excess. A better understanding of 
these processes would then be particularly relevant, if not mandatory.

A neutrino factory is an ideal machine to probe such anomalous decays
of the muon. The pure initial state beam allows to look for
these decays without intrinsic beam contamination. 

\newpage
%
%

%
%

\newpage

\end{document}